\documentclass{paper}

\usepackage{amsmath}
\usepackage{epsfig}

\newcommand{\dd}{\mathrm{d}}

\newcommand{\vettoreErre}{\boldsymbol{r}}

\begin{document}

\title
{Trajectory of a body in a resistant medium: an elementary derivation}

\author{Riccardo Borghi\thanks{\em Dedicated to Franco Gori on his seventy-fifth birthday}\\
Dipartimento di Elettronica Applicata\\
Universit\`a degli Studi ``Roma tre''\\
Via della vasca navale 84, I-00144 Rome, Italy\\
{\tt borghi@uniroma3.it}
}

\maketitle

\begin{abstract}
A didactical exposition of the  classical problem of  the trajectory  determination of a body, subject to the gravity in  a resistant medium, is  proposed. Our revisitation is aimed at showing a derivation of the problem solution  which should be as simple as possible from a technical point of view, in order to be grasped even by first-year undergraduates. A central role in our analysis is played by the so-called ``chain rule'' 
for derivatives, which is systematically used to remove the temporal variable from Newton's law in order to derive the differential equation of  the Cartesian representation of the trajectory, with a considerable reduction of the overall mathematical complexity. In particular, for a resistant medium exerting a force quadratic with respect to the velocity our approach leads to the differential equation of the trajectory, which allows its Taylor series expansion to be derived in an easy way. A numerical comparison of the polynomial approximants obtained by truncating such series with the solution recently proposed through a homotopy analysis is also presented.
\end{abstract}


\section{Introduction}
\label{Sec:Introduction}

The study of the motion of  a body  under the action of the gravity is one of the first applications of the Newton's
laws met by a student in a typical first-year course of general physics. Since the topic is not particularly exciting 
to present,  one is tempted to find unconventional ways to expose the subject in order to render it more  appealing to the 
audience. The determination of the body trajectory, and of the range in particular, is a good candidate for this; 
if the body is in vacuum the laws of motion assume, in the time domain, simple analytical expressions that make the trajectory determination a nearly trivial task. On the other hand, when the body moves in air, where the resistance acting on it is customarily  modeled as a force anti-parallel to its velocity and whose modulus is proportional to the first (for ``slow'' objects) or to the second (for ``fast'' objects) power of the body speed, the mathematical complexity  grows so rapidly that even nowadays the study of the motion in a resistant medium remains an active research field, as witnessed by the recent literature~\cite{Chudinov/2001,Packel/Yuen/2004,Stewart/2006,Yabushita/2007,Vial/2007,Warburton/2010,Stewart/2012,Vial/2012}.

The aim of the present work is to give a didactical presentation of the trajectory determination problem as simple as possible from a technical point of view, in order to be grasped even by first-year undergraduates. We are aware of several nice didactical approaches to this problem appeared in the past~\cite{Purcell/1977,Parker/1977,Lichtenberg/1978,Erlichson/1983,Tan/1987,Groetsch/1996,Groetsch/1997,Price/1998,Timmerman/1999}; nevertheless, we believe that it could be possible to further reduce the mathematical level of the presentation to avoid resorting to concepts, like for instance the hodograph~\cite{Hamilton/1847,Apostolatos/2003}, 
which will become available to the students typically during the second year. 
Our presentation is organized as a sequence of steps of growing complexity: 
starting from the ideal, free-friction case, a resistance linear in the velocity is considered, up to the most difficult (and still open) problem of the motion of a body in the presence of a quadratic resistant force.
For all three cases we propose the same  approach based on the use of Cartesian coordinates, which naturally occur in describing the shape of the body trajectory. Another  aspect we are willing to emphasize  is that the solution of several problems in mechanics often does not require the knowledge of the complete solution of Newton's equations in the time domain, but rather the overall mathematical complexity of an apparent nontrivial problem can be considerably 
reduced by resorting to some ``tricks'' with which the student should become familiar to as soon as possible.\footnote{Richard Feynman was used
to call the collection of such tricks as the ``box of tools'' of any physics student~\cite{Feynamn/Leighton/1985}.}
Among them the use of the so-called ``chain rule'' for derivatives appears to be a technical instrument
of invaluable help to allow nontrivial problems to be addressed also at an elementary level of treatment.\footnote{
A celebrated example of use of the chain rule in mechanics has been provided by Arnold Sommerfeld in deriving the first Kepler's 
law without resorting to the energy conservation~\cite{Sommerfeld/1970}.}
To help the student to familiarize with this technique, the classical parabolic trajectory in vacuum is re-derived
without touching the temporal laws of motion, but rather by removing the temporal variable from the Newton's law, written in the ``natural'' 
Cartesian reference frame, through the use of the chain rule. This is done to introduce in the most 
transparent way to the student the approach subsequently employed when a linear resistance force is added to the dynamical model
and which allows the Cartesian equation of the trajectory to be exactly retrieved only through elementary quadratures.
In the final part of the paper the same methodology is applied to the most realistic case of a quadratic resistance force, for which 
the determination of the analytical expression of the trajectory equation still remains an open problem.
In this case, the proposed approach leads to the differential equation for the trajectory  in Cartesian coordinates,  which, within  the 
small slope approximation, can be integrated in an elementary way to retrieve the classical solution provided by Lamb~\cite{Parker/1977,Lamb/1961}. Moreover, the differential equation turns out to be particularly suitable to be solved by
a power series expansion, whose single terms can, in principle, be evaluated analytically up to arbitrarily high orders.
Finally, the same differential equation is used to provide an elementary derivation of the hodograph of the motion,
thus establishing a sort of changeover toward more advanced approaches.


\section{Trajectory of a body in vacuum}
\label{Sec:MotionInVacuum}

We consider the classical parabolic motion of a point mass $m$ under the action of the sole gravity force, 
so that Newton's second law reduces to
\begin{equation}
\label{Eq:MotionInVacuum.1}
\boldsymbol{a}\,=\,\boldsymbol{g}\,,
\end{equation}
where $\boldsymbol{a}$ denotes the point acceleration and $\boldsymbol{g}$ the 
gravity acceleration. 
\begin{figure}[!ht]
\centerline{\psfig{file=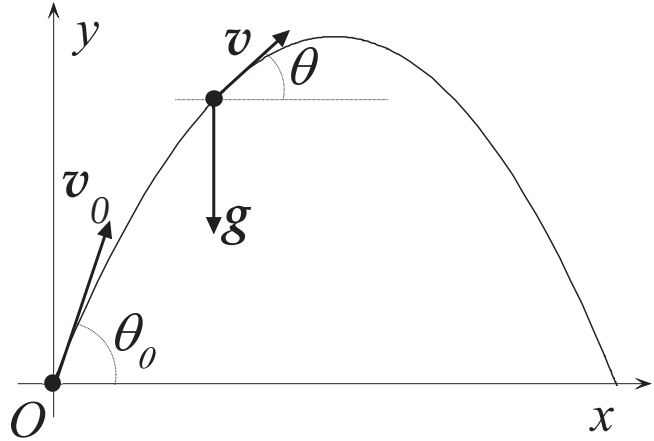,width=7cm,clip=,angle=-90}}
\caption{Geometry of the problem.}
\label{Fig:MotionInVacuum.1}
\end{figure}
In Fig.~\ref{Fig:MotionInVacuum.1} the  Cartesian reference frame $Oxy$ used to describe the motion of 
the point mass is depicted, with $\boldsymbol{v}_0$ denoting the initial velocity (with $x$- and $y$- components
$v_{0,x}$ and $v_{0,y}$, respectively). With respect to
this frame Eq.~(\ref{Eq:MotionInVacuum.1}) splits into the pair of scalar differential equations
\begin{equation}
\label{Eq:MotionInVacuum.2}
\left\{
\begin{array}{l}
\dot v_x\,=\,0\,,\\
\\
\dot v_y\,=\,-g\,,
\end{array}
\right.
\end{equation}
where $v_x$ and $v_y$ denote the $x$- and $y$-component of the velocity, respectively, and
the dot denotes derivation with respect to the time $t$. 
The trajectory of the motion is naturally described, in Cartesian coordinates, by the function $y=y(x)$ and our aim
is to extract the related differential equation without solving Eq.~(\ref{Eq:MotionInVacuum.2}) in the time domain.
In doing so, we have to formally remove from the equations of motion the temporal variable $t$. This can be
done in a very simple way, by using the chain rule for derivates. 

To this end we start from the  relation $\dot{\vettoreErre}\,=\,\boldsymbol{v}$ 
written in $Oxy$, i.e., 
\begin{equation}
\label{Eq:MotionInVacuum.3}
\left\{
\begin{array}{l}
\dot x\,=\,v_x\,,\\
\\
\dot y\,=\,v_y\,,
\end{array}
\right.
\end{equation}
and, after dividing side by side the two equations, we obtain
\begin{equation}
\label{Eq:MotionInVacuum.4}
\begin{array}{l}
\displaystyle
y'(x)\,=\,\frac{\dd y}{\dd x}\,=\,\frac{v_y}{v_x}\,,
\end{array}
\end{equation}
which coincides with  the slope of the trajectory $\eta=\tan\theta$, with $\theta$ being the angle between 
$\boldsymbol{v}$ and the $x$-axis, with the superscript denoting derivative with respect to the spatial
variable.
To find the differential equation for $y(x)$ it is now sufficient to derive both sides 
of Eq.~(\ref{Eq:MotionInVacuum.4}) with respect to $x$ and to take the chain rule, i.e.,
\begin{equation}
\label{Eq:MotionInVacuum.5}
\begin{array}{l}
\displaystyle
\frac{\dd }{\dd x}\,=\,
\frac 1{v_x}\,\frac{\dd }{\dd t}\,,
\end{array}
\end{equation}
into account. 
On substituting from Eq.~(\ref{Eq:MotionInVacuum.5}) into
Eq.~(\ref{Eq:MotionInVacuum.4}) we have
\begin{equation}
\label{Eq:MotionInVacuum.6}
\begin{array}{l}
\displaystyle
\frac{\dd^2 y}{\dd x^2}\,=\,
\frac 1{v_x}\,\frac{\dd }{\dd t}\left(\frac{v_y}{v_x}\right)\,=\,
\frac {\dot{v}_y\,v_x\,-\,\dot{v}_x\,v_y}{v^3_x}\,,
\end{array}
\end{equation}
which, on taking Eq.~(\ref{Eq:MotionInVacuum.2}) into account, leads to
\begin{equation}
\label{Eq:MotionInVacuum.7}
\begin{array}{l}
\displaystyle
\frac{\dd^2 y}{\dd x^2}\,=\,-\frac{g}{v^2_x}\,=\,-\frac{g}{v^2_{0,x}}\,,
\end{array}
\end{equation}
and where, in the last passage, use has been made of the fact that the $x$-component of the velocity
is constant. Equation~(\ref{Eq:MotionInVacuum.7}) can be solved via elementary quadratures; 
in particular, on imposing the ``initial'' (i.e., at $x=0$) conditions
\begin{equation}
\label{Eq:MotionInVacuum.11}
\begin{array}{l}
\displaystyle
y(0)\,=\,0\,,\\
\\
\displaystyle
y'(0)\,=\,\tan\theta_0\,,\\
\end{array}
\end{equation}
the well known equation for the parabolic projectile trajectory is obtained: 
\begin{equation}
\label{Eq:MotionInVacuum.12}
\begin{array}{l}
\displaystyle
y(x)\,=\,x\,\tan\theta_0\,-\,\frac{g}{2v^2_{0,x}}\,x^2\,.
\end{array}
\end{equation}

\section{Trajectory in the presence of a linear resistance force}
\label{Sec:MotionInNewtonianMedium}

Consider now the motion of the point mass in the presence of a 
resistant force proportional to the point velocity, i.e., of the form
$-b\,\boldsymbol{v}$, with $b$ denoting a suitable parameter.
In this case Newton's law can be cast in the following form:
\begin{equation}
\label{Eq:MotionInNewtonianMedium.1}
\boldsymbol{a}\,=\,\boldsymbol{g}\,-\,\frac 1\tau\,\boldsymbol{v}\,,
\end{equation}
where the parameter $\tau=m/b$ rules the temporal scale of the resulting dynamics. 
This, in particular, suggests the use of ``natural'' and dimensionless normalized quantities: $\tau$ will become the temporal unity, the product $g\tau$ (the so-called limit speed) the speed unity, and $g\tau^2$ the displacement unity.\footnote{This is equivalent to formally set $\tau=1$ and $g=1$, in Eq.~(\ref{Eq:MotionInNewtonianMedium.1}).} 
With such units the subsequent equations will be expressed in a dimensionless form, 
by introducing the following scaled time, velocity, and displacement variables:
\begin{equation}
\label{Eq:MotionInNewtonianMedium.11}
\begin{array}{l}
\displaystyle
T\,=\,\frac t{\tau}\,,\\
\\
\displaystyle
(V_x,V_y)\,=\, \frac 1{g\,\tau}\,(v_x,v_y)\,,\\
\\
\displaystyle
(X,Y)\,=\,\frac 1{g\,\tau^2}\,(x,y)\,,
\end{array}
\end{equation}
so that Eq.~(\ref{Eq:MotionInNewtonianMedium.1}) leads to the dimensionless system
\begin{equation}
\label{Eq:MotionInNewtonianMedium.2}
\left\{
\begin{array}{l}
\dot V_x\,=\,-V_x\,,\\
\\
\dot V_y\,=\,-1\,-\,V_y\,,
\end{array}
\right.
\end{equation}
which, once inserted into the dimensionless counterpart of Eq.~(\ref{Eq:MotionInVacuum.6}),
i.e.,
\begin{equation}
\label{Eq:MotionInVacuum.6.Dimensionless}
\begin{array}{l}
\displaystyle
\frac{\dd^2 Y}{\dd X^2}\,=\,
\frac 1{V_x}\,\frac{\dd }{\dd T}\left(\frac{V_y}{V_x}\right)\,=\,
\frac {\dot{V}_y\,V_x\,-\,\dot{V}_x\,V_y}{V^3_x}\,,
\end{array}
\end{equation}
leads to the following differential equation for the trajectory:
\begin{equation}
\label{Eq:MotionInNewtonianMedium.3}
\begin{array}{l}
\displaystyle
\frac{\dd^2 Y}{\dd X^2}\,=\,
\frac {(-1-V_y)\,V_x\,+\,{V}_x\,V_y}{V^3_x}\,=\,-\frac{1}{V^2_x}\,,
\end{array}
\end{equation}
formally identical to Eq.~(\ref{Eq:MotionInVacuum.7}). However, differently from
the previous case, the $x$-component of the velocity is no longer constant, and 
we have to find the dependance of $V_x$ on the variable $X$. This can be easily done on using the 
first row of Eq.~(\ref{Eq:MotionInNewtonianMedium.2}) together with the chain rule 
in Eq.~(\ref{Eq:MotionInVacuum.5}) to remove $T$, thus obtaining
\begin{equation}
\label{Eq:MotionInNewtonianMedium.4}
\begin{array}{l}
\displaystyle
\frac{\dd V_x}{\dd X}\,=\,-1\,,
\end{array}
\end{equation}
which gives
\begin{equation}
\label{Eq:MotionInNewtonianMedium.5}
\begin{array}{l}
V_x\,=\,C\,-\,X\,,
\end{array}
\end{equation}
where the constant $C$ can be found by imposing the initial
condition on $V_x$, i.e., $V_x(0)=V_{0,x}$, so that 
\begin{equation}
\label{Eq:MotionInNewtonianMedium.6}
\begin{array}{l}
V_x(X)\,=\,V_{0,x}\,-\,X\,.
\end{array}
\end{equation}
On substituting from Eq.~(\ref{Eq:MotionInNewtonianMedium.6}) into Eq.~(\ref{Eq:MotionInNewtonianMedium.3}) we obtain 
\begin{equation}
\label{Eq:MotionInNewtonianMedium.7}
\begin{array}{l}
\displaystyle
\frac{\dd^2 Y}{\dd X^2}\,=\,
-\,\frac 1{(V_{0,x}\,-\,X)^2}\,,
\end{array}
\end{equation}
that  can be  solved by using elementary quadratures. In particular,
on taking the initial conditions in Eq.~(\ref{Eq:MotionInVacuum.11}) into account, after some algebra we have
\begin{equation}
\label{Eq:MotionInNewtonianMedium.10}
\begin{array}{l}
\displaystyle
Y\,=\,X\,\tan\theta_0\,+\,\frac X{V_{0,x}}\,+\,\log\left(1-\frac X{V_{0,x}}\right)\,,
\end{array}
\end{equation}
which, on using Eq.~(\ref{Eq:MotionInNewtonianMedium.11}) in order to remove the normalization and to come back to physical quantities, becomes
\begin{equation}
\label{Eq:MotionInNewtonianMedium.12}
\begin{array}{l}
\displaystyle
y(x)\,=\,x\,\tan\theta\,+\,
g\tau^2\,\left[
\frac x{v_{0,x}\tau}\,+\,
\log\left(1-\frac x{v_{0,x}\tau}\right)
\right]\,.
\end{array}
\end{equation}
It should be noted that Eq.~(\ref{Eq:MotionInNewtonianMedium.12}) is known from the literature, although its derivation is customarily carried out by first solving the time-domain dynamical equation~(\ref{Eq:MotionInNewtonianMedium.1}) and only after removing the variable $t$ among the functions $x=x(t)$ and $y=y(t)$.\footnote{See for
instance Refs.~\cite{Stewart/2012,Groetsch/1999}.} 
It is also worth showing to the student how, for small values of drag force, the shape of the trajectory 
tends to become identical to the ideal free-friction case of Eq.~(\ref{Eq:MotionInVacuum.12}).
To this end it is sufficient to consider the limiting expression, for $\tau\to\infty$, of the logarithmic 
function in Eq.~(\ref{Eq:MotionInNewtonianMedium.12}) which, by using a second-order Taylor expansion, 
turns out to be 
\begin{equation}
\label{Eq:MotionInNewtonianMedium.13}
\begin{array}{l}
\displaystyle
\xi\,+\,\log(1-\xi)\,\sim\,-\,\frac{\xi^2}2\,,
\qquad\,\xi \to 0\,.
\end{array}
\end{equation}

\section{Trajectory with a quadratic-speed drag force}
\label{Sec:MotionQuadraticDragForce}

\subsection{Derivation of the differential equation}
\label{SubSec:MotionQuadraticDragForceEQ}

When the resistant medium exerts on the body a force proportional to the 
squared modulus of the speed, Newton's law takes on the following form:\footnote{It could be worth putting into evidence
the fact that, as reported for instance in Ref.~\cite{Groetsch/1999}, Newton himself was aware that a linear resistance model 
was not the most suitable from a physical viewpoint, but rather the resistance offered by media without rigidity would have been expected 
to be proportional to the square of speed, as he wrote at the end of Sec.~I of 
Book~II of his \emph{Principia}~\cite{Newton/1723}:
\begin{quotation}
{\em
C{\ae}terum resistentiam corporum esse in ratione velocitatis, Hypothesis
est magis Mathematica quam Naturalis. Obtinet h{\ae}c ratio quamproxime ubi 
corpora in Mediis rigore aliquo pr{\ae}ditis tardissime moventur. 
In Mediis autem qu{\ae}  rigore omni vacant resistenti{\ae} corporum sunt in duplicata
ratione velocitatum.
}
\end{quotation}
} 
\begin{equation}
\label{Eq:MotionQuadraticDragForce.1}
\boldsymbol{a}\,=\,\boldsymbol{g}\,-\,\frac 1\ell\,v\,\boldsymbol{v}\,,
\end{equation}
where $\ell$ is a parameter whose physical dimensions are those of a length.
Similarly as done for the linear resistant medium, it is worth 
giving Newton's law a dimensionless dress by introducing the following 
scaled variables: 
\begin{equation}
\label{Eq:MotionQuadraticDragForce.1.1}
\begin{array}{l}
\displaystyle
T\,=\,t\,\sqrt{\frac g\ell}\,,\\
\\
\displaystyle
(V_x,V_y)\,=\, \frac 1{\sqrt{g\ell}}\,(v_x,v_y)\,,\\
\\
\displaystyle
(X,Y)\,=\,\frac 1{\ell}\,(x,y)\,,
\end{array}
\end{equation}
so that Eq.~(\ref{Eq:MotionQuadraticDragForce.1}) takes on the form~\cite{Parker/1977}
\begin{equation}
\label{Eq:MotionQuadraticDragForce.2}
\left\{
\begin{array}{l}
\dot V_x\,=\,-V\,V_x\,,\\
\\
\dot V_y\,=\,-1\,-\,V\,V_y\,,
\end{array}
\right.
\end{equation}
where the speed $V$, defined as 
\begin{equation}
\label{Eq:MotionQuadraticDragForce.3}
\begin{array}{l}
V\,=\,\sqrt{V^2_x\,+\,V^2_y}\,,
\end{array}
\end{equation}
acts as a coupling factor. 
On substituting from Eq.~(\ref{Eq:MotionQuadraticDragForce.2}) into 
Eq.~(\ref{Eq:MotionInVacuum.6.Dimensionless}) we find again
\begin{equation}
\label{Eq:MotionQuadraticDragForce.4}
\begin{array}{l}
\displaystyle
\frac{\dd^2 Y}{\dd X^2}\,=\,
\frac {(-1-V\,V_y)\,V_x\,+\,V\,{V}_x\,V_y}{V^3_x}\,=\,-\frac{1}{V^2_x}\,,
\end{array}
\end{equation}
but now it is no longer possible to 
achieve an explicit $X$-dependance for $V_x$, a fact which prevents 
Eq.~(\ref{Eq:MotionQuadraticDragForce.4}) to be solvable via 
simple quadratures. In fact, on applying the chain rule to the first row of 
Eq.~(\ref{Eq:MotionQuadraticDragForce.2}) we have
\begin{equation}
\label{Eq:MotionQuadraticDragForce.4.0}
\begin{array}{l}
\displaystyle
\frac{\dd V_x}{\dd T}\,=\,
V_x\,\frac{\dd V_x}{\dd X}\,=\,
\,-V_xV\,=\,-V_x\,\sqrt{V^2_x\,+\,V^2_y}\,,
\end{array}
\end{equation}
so that
\begin{equation}
\label{Eq:MotionQuadraticDragForce.4.1}
\begin{array}{l}
\displaystyle
\frac{\dd V_x}{\dd X}\,=\,
\,-\sqrt{V^2_x\,+\,V^2_y}\,.
\end{array}
\end{equation}
Moreover, on deriving both sides of Eq.~(\ref{Eq:MotionQuadraticDragForce.4}) with respect to $X$, 
on taking Eq.~(\ref{Eq:MotionQuadraticDragForce.4.1}) into account,  so that
\begin{equation}
\label{Eq:MotionQuadraticDragForce.5}
\begin{array}{l}
\displaystyle
\frac{\dd^3 Y}{\dd X^3}\,=\,
\frac{2}{V^3_x}\,\frac{\dd V_x}{\dd X}\,=\,
-\frac 2{V^2_x}\,\sqrt{1\,+\,\left(\frac{V_y}{V_x}\right)^2}\,=\,
-\frac 2{V^2_x}\,\sqrt{1\,+\,\left(\frac{\dd Y}{\dd X}\right)^2}\,,
\end{array}
\end{equation}
and on using again Eq.~(\ref{Eq:MotionQuadraticDragForce.4})
we obtain
\begin{equation}
\label{Eq:MotionQuadraticDragForce.6}
\begin{array}{l}
\displaystyle
\frac{\dd^3 Y}{\dd X^3}\,=\,
2\,\frac{\dd^2 Y}{\dd X^2}\,\sqrt{1\,+\,\left(\frac{\dd Y}{\dd X}\right)^2}\,,
\end{array}
\end{equation}
which is the differential equation satisfied by the Cartesian representation of the body trajectory $Y=Y(X)$. It has to be solved together with the
initial conditions
\begin{equation}
\label{Eq:MotionQuadraticDragForce.6.1}
\begin{array}{l}
\displaystyle
Y(0)\,=\,0\,,\\
\\
\displaystyle
Y'(0)\,=\,\tan\theta_0\,,\\
\\
\displaystyle
Y''(0)\,=\,-\frac 1{V^2_{0,x}}\,,
\end{array}
\end{equation}
but unfortunately the solution of the Cauchy problem in Eqs.~(\ref{Eq:MotionQuadraticDragForce.6}) and~(\ref{Eq:MotionQuadraticDragForce.6.1})
cannot be given a closed form; nevertheless, nowadays it is easily achievable numerically, due to the availability of several
commercial computational platforms.
 
From a didactical point of view it could be worth showing how Eq.~(\ref{Eq:MotionQuadraticDragForce.6}) leads
in elementary way to a well known analytical approximation of $Y(X)$, called by Parker the ``short-time approximation''~\cite{Parker/1977},\footnote{As the name suggests, the derivation of Parker was done in the time domain. A different derivation can also be found in the beautiful book by  Lamb~\cite{Lamb/1961}.} 
which is valid in the limit of small velocity slopes. In fact, within such approximation it is sufficient to neglect the squared term into the square root 
of Eq.~(\ref{Eq:MotionQuadraticDragForce.6}), in such a way that the differential equation reduces to
\begin{equation}
\label{Eq:MotionQuadraticDragForce.6.2}
\begin{array}{l}
\displaystyle
\frac{\dd^3 Y}{\dd X^3}\,\simeq\,2\,\frac{\dd^2 Y}{\dd X^2}\,,
\end{array}
\end{equation}
which can be  solved simply by iterated elementary quadratures. In particular, on taking Eq.~(\ref{Eq:MotionQuadraticDragForce.6.1}) into account,
after simple algebra we obtain the approximated solution, say $Y_{\rm st}$, as
\begin{equation}
\label{Eq:MotionQuadraticDragForce.6.3}
\begin{array}{l}
\displaystyle
Y_{\rm st}\,=\,\frac{1-\exp(2X)}{4\,V^2_{0,x}}\,+\,X\,\left(\frac 1{2V^2_{0,x}}\,+\,\tan\theta_0\right)\,,
\end{array}
\end{equation}
which coincides with the equation given at the end of Sec.~III in Ref.~\cite{Parker/1977} and with Eq.~(26) at p.~297 of Ref.~\cite{Lamb/1961}.

\subsection{Power series expansion of the trajectory}
\label{SubSec:PowerSeries}

It must be appreciated that the mathematical structure of 
Eq.~(\ref{Eq:MotionQuadraticDragForce.6}) turns out to be particularly suitable 
to derive in a simple way the Taylor series expansion of the trajectory, namely
\begin{equation}
\label{Eq:MotionQuadraticDragForce.6.4}
\begin{array}{l}
\displaystyle
Y(x)\,=\,\sum_{k=0}^\infty\,\frac{a_k}{k!}\,X^k\,,
\end{array}
\end{equation}
where $a_k\,=\,Y^{(k)}(0)$ denotes the $k$th-order spatial derivative of $Y(X)$, evaluated at $X=0$. In fact,  starting from the initial conditions
given in Eq.~(\ref{Eq:MotionQuadraticDragForce.6.1}) we have for the coefficient $a_3$ the expression
\begin{equation}
\label{Eq:MotionQuadraticDragForce.6.5}
\begin{array}{l}
\displaystyle
a_3\,=\,Y'''(0)\,=\,2\,Y''(0)\,\sqrt{1+[Y'(0)]^2}\,=\,
-\frac 2{V^2_{0,x}\,\cos\theta_0}\,,
\end{array}
\end{equation}
which differs from the corresponding coefficient in the power series expansion of 
$Y_{\rm st}$ by a factor $\cos\theta_0$ that, in the case of small slopes,
can be replaced by the unity. As far as the coefficient $a_4$ is concerned, it is
sufficient to derive both sides of Eq.~(\ref{Eq:MotionQuadraticDragForce.6}) with respect to $X$ to obtain
\begin{equation}
\label{Eq:MotionQuadraticDragForce.6.6}
\begin{array}{l}
\displaystyle
Y^{(4)}\,=\,2\,\frac{Y'''\,(1+Y^{\prime\,2})\,+\,Y'\,Y''}{\sqrt{1+Y^{\prime\,2}}}\,,
\end{array}
\end{equation}
which, after rearranging and simplifying, gives at once
\begin{equation}
\label{Eq:MotionQuadraticDragForce.6.7}
\begin{array}{l}
\displaystyle
a_{4}\,=\,-\frac 4{V^4_{0,x}}\,\left(V^2_0\,-\,\frac{\sin\theta_0}2\right)\,,
\end{array}
\end{equation}
whereas the short-time approximation would provide
\begin{equation}
\label{Eq:MotionQuadraticDragForce.6.7.1}
\begin{array}{l}
\displaystyle
Y^{(4)}_{\rm st}(0)\,=\,-\frac 4{V^2_{0,x}}\,.
\end{array}
\end{equation}
It is clear that, in principle, all terms of the power series expansion in Eq.~(\ref{Eq:MotionQuadraticDragForce.6.4})
could be evaluated by iterating this approach; however, on increasing the truncation of the expansion
the complexity of the analytical expressions of the coefficients $a_k$ rapidly grows,\footnote{As a further example we give the  expression
of the 5th-order coefficient:
\[
a_5\,=\,-\frac{8\,\cos\theta_0}{V^6_{0,x}}\,\left(V^4_0\,-\,2\,V^2_0\,\sin\theta_0\,+\,\frac 14\,\cos^2\theta_0\right)\,,
\]
which should be compared to its ``short-time'' version, given by $-8/V^2_{0,x}$.
}
although it is easy to implement their evaluation, up to arbitrarily high orders, by using any commercial symbolic computational platform.

To give an idea of the performances of the above polynomial approximation, we analyze a case already studied in the past, related to
the motion of a volleyball in air~\cite{Yabushita/2007}.
\begin{figure}[!ht]
\centerline{\psfig{file=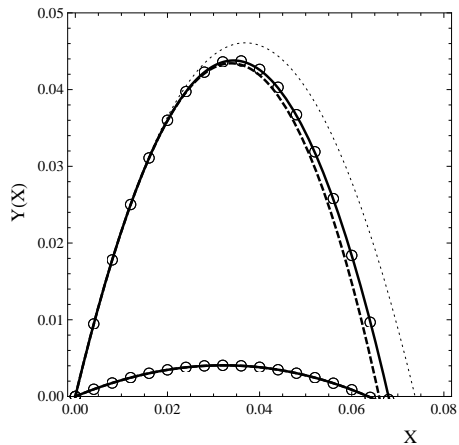,width=6cm,clip=,angle=0}}
\caption{Trajectory (within dimensionless spatial variables) of a volleyball in air 
($\ell\,\simeq\,27.64$~m) as numerically evaluated from Eq.~(\ref{Eq:MotionQuadraticDragForce.6}) (open circles) together with
the  3rd-order (dashed curve) and the 4th-order (solid curve)  polynomial approximation obtained by truncation of the power series 
expansion in Eq.~(\ref{Eq:MotionQuadraticDragForce.6.4}). The upper curve refer to the initial conditions $(v_{0,x},v_{0,y})=(2,5)$~m/s, while the other curve to the initial conditions $(v_{0,x},v_{0,y})=(6,3/2)$~m/s, in order to be comparable to Fig.~4 of Ref.~\cite{Yabushita/2007}. The dotted curve represents the trajectory corresponding to the ideal case in vacuum.}
\label{Fig:MotionQuadraticDragForce.1}
\end{figure}
Figure~\ref{Fig:MotionQuadraticDragForce.1} shows the ball trajectory (within dimensionless spatial variables) for an initial velocity with components $(v_{0,x},v_{0,y})=(2,5)$~m/s
in a medium characterized by $\ell\,\simeq\,27.64$~m. Such data refer to the results shown in Fig.~4 of Ref~\cite{Yabushita/2007}.
The exact solution, obtained by solving numerically Eq.~(\ref{Eq:MotionQuadraticDragForce.6}) through the standard command {\tt NDSolve} of the symbolic language {\tt Mathematica} is represented by the circles, while the dashed and the solid curves are the polynomial
approximations obtained by truncating the series in Eq.~(\ref{Eq:MotionQuadraticDragForce.6.4}) up to $k=3$ and $k=4$, respectively.
The dotted curve represents the trajectory corresponding to the ideal case in vacuum.
From the figure it is  seen that the fourth-order polynomial displays an excellent  agreement with the exact curve along the whole range of interest of $x$.
To make the comparison between our figure and Fig.~4 of Ref.~\cite{Yabushita/2007} more complete, the trajectory corresponding to
the initial condition $(v_{0,x},v_{0,y})=(6,3/2)$~m/s  is also plotted. In this case the 3rd- and the 4th-order approximants are practically undistinguishable.

\begin{figure}[!ht]
\centerline{\psfig{file=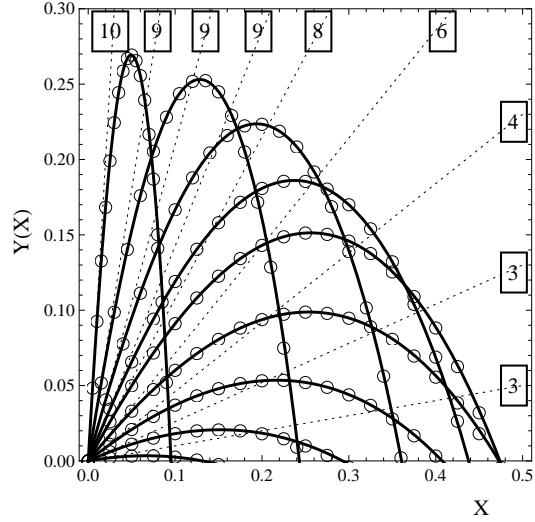,width=7cm,clip=,angle=0}}
\caption{The same as in Fig.~\ref{Fig:MotionQuadraticDragForce.1}, but for 
$v_0\,=\,14$~m/s and for several values of the initial angle $\theta_0$, extracted from Fig.~6 of Ref.~\cite{Yabushita/2007}. 
for each curve it is shown the exact solution (open circles) together with an $N$th-order polynomial approximation (solid curve), 
with $N$ (labelled for each curve) being the truncation order of the power series which has been chosen in order to allows a satisfactory agreement 
 to be achieved.}
\label{Fig:MotionQuadraticDragForce.2}
\end{figure}

Figure~\ref{Fig:MotionQuadraticDragForce.2} shows the results corresponding to $v_0\,=\,14$~m/s and to several values of the initial angle
$\theta_0$, retrieved from Fig.~6 of Ref.~\cite{Yabushita/2007}. In particular, for each curve it is shown the exact solution (open circles)
together with an $N$th-order polynomial approximation (solid curve), with $N$ (labelled for each curve) being the lowest value of the truncation order of the power series which
guarantees a satisfactory agreement (at least at a visual level) to be achieved.

\subsection{An elementary derivation of the hodograph}
\label{SubSec:Hodograph}

In the present section we want to show how the third-order differential equation~(\ref{Eq:MotionQuadraticDragForce.6})
leads in a simple way to the hodograph of the motion~\cite{Hamilton/1847,Apostolatos/2003}.
To this end we first recast Eq.~(\ref{Eq:MotionQuadraticDragForce.6}) in terms of the 
slope $\eta=Y'$ as follows:
\begin{equation}
\label{Eq:MotionQuadraticDragForce.7}
\begin{array}{l}
\displaystyle
\frac{\dd^2 \eta}{\dd X^2}\,=\,
2\,\frac{\dd \eta}{\dd X}\,\sqrt{1\,+\,\eta^2}\,,
\end{array}
\end{equation}
which, on letting
\begin{equation}
\label{Eq:MotionQuadraticDragForce.7.1}
\frac{\dd \eta}{\dd X}\,=\,\eta'\,,
\end{equation}
becomes
\begin{equation}
\label{Eq:MotionQuadraticDragForce.8}
\frac{\dd \eta'}{\dd X}\,=\,2\,\eta'\,\sqrt{1\,+\,\eta^2}\,.
\end{equation}
Although the differential equation cannot be solved in analytical terms, it is still possible to find an explicit functional expression for the quantity $\eta'$ whether it is thought of as a \emph{function of $\eta$}. Again the trick is to use the chain rule, by letting 
\begin{equation}
\label{Eq:MotionQuadraticDragForce.9}
\begin{array}{l}
\displaystyle
\frac{\dd }{\dd X}\,=\,
\frac{\dd \eta}{\dd X}\,
\frac{\dd }{\dd \eta}\,=\,
\eta'\,\frac{\dd }{\dd \eta}\,,
\end{array}
\end{equation}
so that Eq.~(\ref{Eq:MotionQuadraticDragForce.8}) becomes
\begin{equation}
\label{Eq:MotionQuadraticDragForce.9.1}
\begin{array}{l}
\displaystyle
\frac{\dd \eta'}{\dd \eta}\,=\,2\,\sqrt{1\,+\,\eta^2}\,,
\end{array}
\end{equation}
and $\eta'$ can be retrieved through a simple quadrature, which gives
\begin{equation}
\label{Eq:MotionQuadraticDragForce.10}
\begin{array}{lcl}
\displaystyle
\eta'&=&\eta\,\sqrt{1\,+\,\eta^2}\,+\,\mathrm{arcsinh} \eta\,+\,C\,=\,\\
&&\\
&=&
\eta\,\sqrt{1\,+\,\eta^2}\,+\,\log(\eta\,+\,\sqrt{1\,+\,\eta^2})\,+\,C\,.
\end{array}
\end{equation}
The constant $C$ has to be determined by imposing that at the start of the trajectory, when its slope
is $\eta_0=\tan\theta_0$, the corresponding value of $\eta'$, say $\eta'_0$ is, according to 
Eq.~(\ref{Eq:MotionQuadraticDragForce.4}),
\begin{equation}
\label{Eq:MotionQuadraticDragForce.11}
\begin{array}{lcl}
\displaystyle
\eta'_0\,=\,-\frac 1{V^2_{0,x}}\,=\,-\frac 1{V^2_0\,\cos^2\theta_0}\,,
\end{array}
\end{equation}
so that, after simple algebra, we obtain
\begin{equation}
\label{Eq:MotionQuadraticDragForce.12}
\begin{array}{l}
\displaystyle
C\,=\,-\frac 1{V^2_{0,x}}\,-\,f(\theta_0)\,,
\end{array}
\end{equation}
where the function $f(\cdot)$ is defined by
\begin{equation}
\label{Eq:MotionQuadraticDragForce.13}
\begin{array}{l}
\displaystyle
f(\theta)\,=\,\frac{\sin\theta}{\cos^2\theta}\,+\,\log\frac{1+\sin\theta}{\cos\theta}\,.
\end{array}
\end{equation}
On substituting from Eqs.~(\ref{Eq:MotionQuadraticDragForce.12}) and~(\ref{Eq:MotionQuadraticDragForce.13}) into
Eq.~(\ref{Eq:MotionQuadraticDragForce.10}) the following differential equation for the function $\eta(X)$ is then obtained:
\begin{equation}
\label{Eq:MotionQuadraticDragForce.14}
\begin{array}{l}
\displaystyle
\frac{\dd \eta}{\dd X}\,=\,-\frac 1{V^2_{0,x}}\,+\,
\eta\,\sqrt{1\,+\,\eta^2}\,+\,\log(\eta\,+\,\sqrt{1\,+\,\eta^2})\,-\,f(\theta_0)\,,
\end{array}
\end{equation}
which, on recalling that $\eta=\tan\theta$ and that the l.h.s. does coincide with $Y''(X)$, after algebra 
can be recast as follows:
\begin{equation}
\label{Eq:MotionQuadraticDragForce.15}
\begin{array}{l}
\displaystyle
\frac{\dd^2 Y}{\dd X^2}\,=\,-\frac 1{V^2_{0,x}}\,+\,f(\theta)\,-\,f(\theta_0)\,,
\end{array}
\end{equation}
and, on taking Eq.~(\ref{Eq:MotionQuadraticDragForce.4}) into account, after rearranging leads to the well known expression of the hodograph~\cite{Timoshenko/1948}
\begin{equation}
\label{Eq:MotionQuadraticDragForce.16}
\begin{array}{l}
\displaystyle
V\,=\,\frac{V_{0,x}}{\cos\theta\,\sqrt{1+V^2_{0,x}\,[f(\theta_0)\,-\,f(\theta)]}}\,.
\end{array}
\end{equation}

\section{Conclusions}
\label{Sec:Conclusions}

An elementary exposition of the trajectory  determination problem for a body launched at an arbitrary direction and subject to the gravity in the presence of a resistant medium has been proposed. The main task of such exposition was to limit at most the mathematical complexity in order to be appreciable also by first-year undergraduate students, especially in the case of a quadratic resistant medium. To this end, the use of time domain  Newton's laws of the motion was systematically avoided by removing, through the use of the  derivative `chain rule', the temporal variable before solving them. In this way the differential equation for the representation of the trajectory has been derived in a Cartesian reference frame which turns out to be naturally suited to represent the trajectory shape. In the study of the motion in vacuum and in a resistant medium linear in the velocity the proposed approach was able to retrieve, in an elementary way, the well known exact expressions of 
the trajectory equation. For a  body moving in a resistant medium which exerts a force quadratic with respect to
the velocity, the proposed approach allowed  the differential equation for the Cartesian representation 
of the trajectory to be derived in a rather simple way. From the same differential equation simple polynomial approximants
of the trajectory, having the form of truncated power series expansions, have been built up to arbitrary values 
of the truncation order and compared to the results recently obtained through a homotopy analysis based approach.
Finally, to complete such a sort of changeover toward more advanced mathematical treatments of the problem, 
our  approach was also used to derive, again in an elementary way, the closed form of the motion hodograph, 
a well known result from the literature. 

\section*{Acknowledgment}

I wish to thank Turi Maria Spinozzi for his invaluable help during the preparation of the manuscript.



\newpage

\begin{thebibliography}{00}

\bibitem{Chudinov/2001} P. S. Chudinov,
``The motion of a point mass in a medium with a square law of drag,''
J. Appl. Math. Mech. \textbf{65,} 421 - 426 (2001).

\bibitem{Chudinov/2004} P. S. Chudinov,
``Analytical investigation of point mass motion in midair,''
Eur. J. Phys. \textbf{25,} 73 - 79 (2004). 

\bibitem{Packel/Yuen/2004}
E. W. Packel and D. S. Yuen,
``Pojectile motion with resistance and the Lambert W function,"
The College Mathematical Journal \textbf{35,} 337 - 350 (2004).

\bibitem{Stewart/2006} S. M. Stewart, 
``An analytic approach to projectile motion in a linear resisting medium,''
Int. J. Math. Educ. Sci. Technol. \textbf{37,} 411 - 431 (2006).

\bibitem{Yabushita/2007} K. Yabushita, M. Yamashita,  and K. Tsuboi,
``An analytic solution of projectile motion with the
quadratic resistance law using the homotopy analysis
method,'' 
J. Phys. A: Math. Theor. \textbf{40,}  8403 - 8416 (2007).

\bibitem{Vial/2007} A. Vial,
``Horizontal distance travelled by a mobile experiencing a quadratic drag
force: normalized distance and parametrization,''
Eur. J. Phys. \textbf{28,} 657 - 663 (2007). 

\bibitem{Warburton/2010} R. D. H. Warburton, J. Wang, J. Burgd\"orfer
``Analytic Approximations of Projectile Motion with
Quadratic Air Resistance,''
J. Service Science \& Management \textbf{3,} 98 - 105 (2010).

\bibitem{Stewart/2012} S. M. Stewart, 
``On the trajectories of projectiles depicted in early ballistic woodcuts,''
Eur. J. Phys \textbf{33,}  149 - 66 (2012).

\bibitem{Vial/2012} A. Vial,
``Fall with linear drag and Wien's displacement law: approximate solution and Lambert function,''
Eur. J. Phys. \textbf{33,} 751 - 755 (2012). 

\bibitem{Purcell/1977} E. M. Purcell,
``Life at low Reynolds number,''
Am. J. Phys. \textbf{45,} 3 - 11 (1977).

\bibitem{Parker/1977} G. W. Parker,
``Projectile motion with air resistance quadratic in the speed,''
Am. J. Phys. \textbf{45,} 606 - 610 (1977).

\bibitem{Lichtenberg/1978} D. B. Lichtenberg and J. G. Wills,
``Maximizing the range of the shot put,''
Am. J. Phys. \textbf{46,} 546 - 549 (1978).

\bibitem{Erlichson/1983} H. Erlichson,
``Maximum projectile range with drag and lift, with particular application to golf,''
Am. J. Phys. \textbf{51,} 357 - 362 (1983).

\bibitem{Tan/1987} A. Tan, C. H. Frick, and O. Castillo
``The fly ball trajectory: An older approach revisited,''
Am. J. Phys. \textbf{55,} 37 - 40 (1987).

\bibitem{Groetsch/1996} C. W. Groetsch,
``Tartaglia's Inverse Problem in a Resistive Medium,''
The American Mathematical Monthly \textbf{103,} 546 - 551 (1996). 

\bibitem{Groetsch/1997} C. W. Groetsch,
``On the optimal angle of projection in general media,''
Am. J. Phys. \textbf{65,} 797 - 799 (1997).

\bibitem{Price/1998} R. H. Price and J. D. Romano,
``Aim high and go far — Optimal projectile launch angles greater than 45$^\circ$,''
Am. J. Phys. \textbf{66,} 110 - 113 (1998).

\bibitem{Timmerman/1999}
P. Timmerman and J. P. van der Weele,
``On the rise and fall of a ball with linear or quadratic drag,''
Am. J. Phys. \textbf{67,} 538 - 546 (1999).

\bibitem{Hamilton/1847} W. R. Hamilton,
``The hodograph, or a new method of expressing in symbolical language
the Newtonian law of attraction,''
Proc. Royal Irish Academy \textbf{3,} 344 - 353 (1847).

\bibitem{Apostolatos/2003} T. A. Apostolatos
``Hodograph: A useful geometrical tool for solving some difficult problems in
dynamics,''
Am. J. Phys. \textbf{71,} 261 - 266 (2003).

\bibitem{Feynamn/Leighton/1985} R. P. Feynman and R. Leighton,
{\em Surely you're joking, Mr. Feynman!}
(Norton \& Co., NY, 1985).

\bibitem{Sommerfeld/1970} A. Sommerfeld, 
\emph{Lectures on Theoretical Physics. I. Mechanics}
(Academic Press, 1970)

\bibitem{Lamb/1961} H. Lamb,
{\em Dynamics}
(Cambridge University Press, Cambridge, 1961).

\bibitem{Groetsch/1999} C. W. Groetsch,
{\em Inverse Problems: Activities for Undergraduates}
(The Mathematical Association of America, 1999).

\bibitem{Newton/1723} I. Newton,
{\em Philosophi\ae \, Naturalis Principia Mathematica} (1723).

\bibitem{Timoshenko/1948} S. Timoshenko and D.H. Young,
\emph{Advanced Dynamics} 
(McGrow-Hill, New York, 1948).

\end{thebibliography}
\end{document}